\documentclass{article}
\usepackage{amssymb}
\usepackage{amsfonts}
\usepackage{amsmath}
\usepackage{epsfig}
\usepackage{graphicx}

\begin{document}

\title{Quantum control of quasi-collision states: A protocol for hybrid
fusion}
\author{R. Vilela Mendes\thanks{%
e-mail: rvilela.mendes@gmail.com, rvmendes@fc.ul.pt;
http://label2.ist.utl.pt/vilela/} \\
CMAFCIO, Universidade de Lisboa, C6 - Campo Grande, 1749-016 Lisboa}
\date{ }
\maketitle

\begin{abstract}
When confined to small regions quantum systems exhibit electronic and
structural properties different from their free space behavior. These
properties are of interest, for example, for molecular insertion, hydrogen
storage and the exploration of new pathways for chemical and nuclear
reactions.

Here, a confined three-body problem is studied, with emphasis on the study
of the "quantum scars" associated to dynamical collisions. For the
particular case of nuclear reactions it is proposed that a molecular cage
might simply be used as a confining device with the collision states
accessed by quantum control techniques.
\end{abstract}

\section{Introduction}

\subsection{Confined vs nonconfined systems}

When confined to small space regions molecular systems exhibit electronic
and structural properties different from their free space behavior (\cite%
{confin1} - \cite{confin3}). Among other reasons, the study of confined
molecular systems is of interest in view of recent techniques for the
synthesis of nanostructured materials which could serve as containers for
molecular insertion. Examples are the insertion of molecules into fullerene
cages as well as the hydrate structures for hydrogen storage.

The existence of new pathways for reactions in confinement is another
interesting possibility. This applies both to chemical and nuclear
reactions. For chemical reactions it is obvious that confinement in a small
space enclosure, by itself, enhances the overlap and interaction of electron
orbitals. For nuclear reactions, however, confinement is clearly not
sufficient because of the strong Coulomb barrier. Therefore some other
complementary mechanism must be found to overcome the Coulomb barrier. It is
perhaps useful to remember that also in magnetic confinement fusion, the
magnetic fields only provide confinement and not the nuclear collisions
needed for fusion. There the additional mechanism is microwave heating. For
nuclei confined in a molecular cage, microwave heating in inappropriate as
it would also destroy the confining cage. Therefore a subtler quantum
control mechanism has to be found. The next subsection briefly describes a
situation which might provide such a possibility.

\subsection{Unstable classical orbits and quantum scars}

For some time it was believed that, in systems with ergodic classical
motion, the squared eigenfunctions would coincide, in the semiclassical
limit, with the projection of the microcanonical phase-space measure \cite%
{Shnirelman} \cite{Zelditch} \cite{Colin}. Actually, what the exact results,
that were proved in this context, show is that, for a classically ergodic
quantum system, there is an eigenvalue sequence for the density such that
the corresponding quantum densities converge weakly to the Liouville
measure. Therefore, the observation of states that do not fit these
expectations does not contradict the exact mathematical results. The
convergence may be very slow and nothing forbids the existence of other
subsequences converging to measures different from the Liouville measure.

In fact, wave functions were found which are concentrated near the classical
unstable periodic orbits. When this happens one says that the quantum state
is scarred by the unstable periodic orbit or that one has a \textit{quantum
scar}. Such states have been observed at first in numerical simulations and,
for example, in semiconductor quantum-well tunneling experiments \cite%
{Wilkinson}.

The first theory of scars was proposed by Heller \cite{Heller}, further
developed by a number of authors \cite{Bogomolny} \cite{Berry} \cite%
{Feingold}. By scarring the quantum spectrum, quantum scars are another gift
of quantum mechanics, in the sense that unstable orbit configurations that
are unobservable in a classical situation, become well defined quantum
states which may be practically used by resonant excitation and quantum
control. A particular type of scars are those associated to saddle points of
the potentials. Their existence \cite{Vilela1} and potential applications
have been discussed elsewhere \cite{Vilela2}. They were called \textit{%
saddle scars}. It was pointed out that they might be of interest in the
characterization of collision states in the many-body problem, in particular
when the collision points are classically unstable. One of the simplest, yet
potentially interesting, cases occurs in the 3-body problem when both
attractive and repulsive forces are at play. As an example consider a system
with two positive and one negative charge interacting by Coulomb forces. The
potential is%
\begin{equation}
V\left( R,x,y\right) =Z\left( \frac{Z}{R}-\frac{1}{\sqrt{x^{2}+\left( \frac{R%
}{2}-y\right) ^{2}}}+\frac{1}{\sqrt{x^{2}+\left( \frac{R}{2}+y\right) ^{2}}}%
\right)  \label{1.1}
\end{equation}%
$Z$ being the charge of the positive charges and one the negative charge. $%
R,x,y$ are coordinates in the plane of the three particles with the positive
charges placed symmetrically to the origin (Fig.\ref{plane}).

\begin{figure}[htb]
\centering
\includegraphics[width=0.5\textwidth]{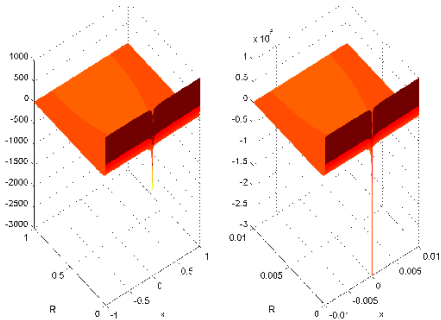}
\caption{Three-particle potential at $y=0$}
\label{potential}
\end{figure}

Fig.\ref{potential} displays the potential in the $\left( R,x\right) -$plane
when $y=0$, at two resolution scales. One sees an attractive singularity at $%
R=x=0$, but the region where the potential is negative is an extremely
narrow one around $x=0$. This singularity is only attractive in the $R$ and $%
x$ directions because for $y\neq 0$ it moves to $R=2\left\vert y\right\vert $%
. Hence this singular point behaves qualitatively like a saddle.

The point $R=x=y=0$ is a collision
point of the three particles. However, because of the unstable nature of
this point and the narrowness of the negative potential region, this
configuration will not be observed in a classical equilibrium setting. Of
course, in a full ergodic chaotic system, confined to a finite volume, there
would be some small occurrence probability. This situation was studied
before \cite{Vilela3}, but the nevertheless non-vanishing quasi-collision
rates that are obtained are too small to be of practical interest. In
addition, to try and induce collisions in a confined system by chaotizing
it, for example, by a temperature increase or a sonic wave not only risks
the destruction of the confinement cage but also, because of the chaotic
nature of the event, leads to basically irreproducible events.

Therefore, for quasi-collisions of many-body systems, involving both
attractive and repulsive forces, to be of practical interest, in chemical or
nuclear situations, it seems better to explore the quantum nature of the
problem, in particular the scar nature of classically unstable quantum
states. And then to address directly these states by quantum control
techniques. A precondition is, of course, to establish the existence of such
states. A first step in this direction was taken in \cite{Vilela4} where a
configuration of two positive charges in a octahedral cage was considered,
the vertices of the cage being occupied by atoms with a partially filled
shell. One-electron energy levels were studied in a basis that contained
both $d$-orbitals centered at the vertices and $s$-orbitals centered at the
positive charges. Although the ground states that are obtained correspond to
large separations of the positive charges, some excited states were found
that have large quasi-collision probabilities. In this paper a similar
situation is studied, with two positive and one negative charge confined in
a cage, but using a much larger state basis. For the diagonalization of the
Hamiltonian, a finite-difference method is used, the size of the basis being
the number of discretization points in the cage. Up to $2.19\times 10^{4}$
basis states were used. Denoting by $R$ the distance between the positively
charged particles and by $z$ all other coordinates and labelling the
eigenstates as $\psi \left( R,z\right) $, the \textit{quasi-collision
probability} is defined as%
\begin{equation}
I_{0}=\int dz\left\vert \psi \left( 0,z\right) \right\vert ^{2}  \label{1.2}
\end{equation}%
As in \cite{Vilela4}, many excited states with $I_{0}\neq 0$ are found.

It must be pointed out that such states which have a scar-like nature can
only be observed in a fully dynamical treatment when $R$ is a dynamical
variable and not an average value obtained by some a-posteriori minimization
problem. In the next section this point is emphasized by exhibiting the
limitations of the quasi-static approximation for the three-body problem.

\section{A charged three-body problem: limitations of the quasi-static
approximation}

\begin{figure}[htb]
\centering
\includegraphics[width=0.5\textwidth]{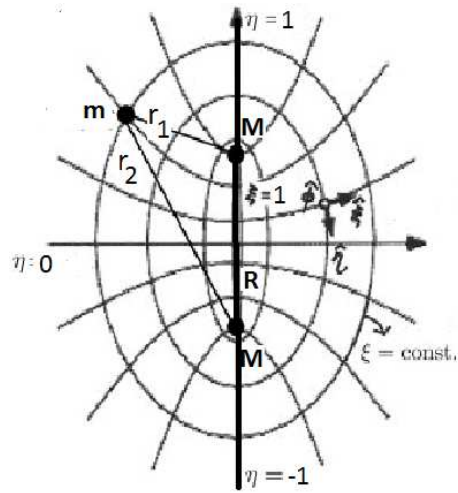}
\caption{Prolate
spheroidal coordinates}
\label{coordinates}
\end{figure}

Here one deals with a Coulomb system of two positively $Z-$charged particles
of mass $M$ interacting with a particle of mass $m$ and unit negative
charge. And, for the moment, one deals with the problem in the full $3-$%
space, not on a confined volume. Define prolate spheroidal coordinates (Fig.%
\ref{coordinates}),

with $1\leq \xi <\infty $ and $-1\leq \eta \leq 1$ being the
spheroidal coordinates in the plane of the three particles and $\phi $ the
angular coordinate of the mass $m$ particle around the $z-$axis defined by
the two mass $M$ particles. In the reference frame of the $z-$axis with the $%
M$ particles placed symmetrically about the origin, $R/2$ is the only
dynamical variable of the mass $M$ particles. In this frame the Coulomb
interaction Hamiltonian is 
\begin{equation}
H=-\frac{\hslash ^{2}}{2M}8\frac{\partial ^{2}}{\partial R^{2}}-\frac{%
\hslash ^{2}}{2m}\Delta _{m,1}+V\left( R,\xi ,\eta \right)  \label{2.1}
\end{equation}%
with%
\begin{equation}
\Delta _{m,1}=\frac{4}{R^{2}\left( \xi ^{2}-\eta ^{2}\right) }\left\{ \frac{%
\partial }{\partial \xi }\left( \xi ^{2}-1\right) \frac{\partial }{\partial
\xi }+\frac{\partial }{\partial \eta }\left( 1-\eta ^{2}\right) \frac{%
\partial }{\partial \eta }+\frac{\partial }{\partial \phi }\left( \frac{\xi
^{2}-\eta ^{2}}{\left( \xi ^{2}-1\right) \left( 1-\eta ^{2}\right) }\right) 
\frac{\partial }{\partial \phi }\right\}  \label{2.2}
\end{equation}%
and%
\begin{equation}
V\left( R,\xi ,\eta \right) =\frac{Ze^{2}}{4\pi \varepsilon _{0}}\left( 
\frac{Z}{R}-\frac{2}{R\left( \xi -\eta \right) }-\frac{2}{R\left( \xi +\eta
\right) }\right)  \label{2.3}
\end{equation}%
$Z$ being the ratio of the charges of the mass $M$ and the mass $m$
particles. With $G^{2}=\frac{mZe^{2}}{2\hslash ^{2}\pi \varepsilon _{0}}$
and $\mu =\frac{m}{M}$ one has%
\begin{equation}
\frac{2m}{\hslash ^{2}}H=-8\mu \frac{\partial ^{2}}{\partial R^{2}}-\Delta
_{m,1}+\frac{G^{2}}{R}\left( Z-\frac{4\xi }{\xi ^{2}-\eta ^{2}}\right)
\label{2.4}
\end{equation}%
In the usual treatments of the ground and excited states of the ionized
hydrogen molecule, $R$ is treated as a parameter to be fixed by minimizing
the energy associated to each $\psi \left( \xi ,\eta \right) $ wave
function. This provides for each state (ground or excited) the mean value of
the coordinate $R$. If, however, one wants information on the
quasi-collision probability of the particles, the important issue is the
value the wave function at $R=0$, hence $R$ should have been treated as a
dynamical variable. Because $R$ enters in the second and third term of Eq.(%
\ref{2.4}) with different powers, a complete separation of variables is not
possible. A partial separation which, although better than a purely static
assumption for the mass $M$ particles, is not accurate is to solve the
following eigenvalue problem for each fixed $R$%
\begin{equation}
\left\{ -\frac{R}{4}\Delta _{m,1}+\frac{G^{2}}{4}\left( Z-\frac{4\xi }{\xi
^{2}-\eta ^{2}}\right) \right\} \psi _{\beta ,\alpha }\left( R,\xi ,\eta
\right) \chi _{\alpha }\left( \phi \right) =\lambda _{\beta ,\alpha }\left(
R\right) \psi _{\beta ,\alpha }\left( R,\xi ,\eta \right) \chi _{\alpha
}\left( \phi \right)  \label{2.5}
\end{equation}%
and then, when for each set of quantum numbers $\beta ,\alpha $ (associated
to the variables $\xi ,\eta $ and $\phi $) the function $\lambda _{\beta
,\alpha }\left( R\right) $ is found, to obtain the $R-$dependence of the
wave function from%
\begin{equation*}
\left\{ -8\mu \frac{\partial ^{2}}{\partial R^{2}}+\frac{4\lambda _{\beta
,\alpha }\left( R\right) }{R}\right\} \psi _{\beta ,\alpha }\left( R,\xi
,\eta \right) \chi _{\alpha }\left( \phi \right) =\frac{2m}{\hslash ^{2}}%
E\psi _{\beta ,\alpha }\left( R,\xi ,\eta \right) \chi _{\alpha }\left( \phi
\right)
\end{equation*}%
$V_{\beta ,\alpha }\left( R\right) =\frac{4\lambda _{\beta ,\alpha }\left(
R\right) }{R}$ being an effective potential for the $R-$dependence of the
wave function. Let $\chi _{\alpha }\left( \phi \right) =\exp \left( i\alpha
\phi \right) $ with $\alpha $ an integer. Now, for each fixed $R$,
separation of the variables $\psi _{\beta ,\alpha }\left( R,\xi ,\eta
\right) =\psi _{\beta ,\alpha }\left( R,\xi \right) \psi _{\beta ,\alpha
}\left( R,\eta \right) $ yields%
\begin{eqnarray}
&&\left\{ \frac{\partial }{\partial \xi }\left( \xi ^{2}-1\right) \frac{%
\partial }{\partial \xi }-\frac{\alpha ^{2}}{\xi ^{2}-1}-R\frac{G^{2}}{4}%
\left( Z\xi ^{2}-4\xi \right) -\Lambda _{\alpha ,R}\right\} \psi _{\beta
,\alpha }\left( R,\xi \right) =-R\lambda _{\beta ,\alpha }\left( R\right)
\xi ^{2}\psi _{\beta ,\alpha }\left( R,\xi \right)  \notag \\
&&\left\{ \frac{\partial }{\partial \eta }\left( 1-\eta ^{2}\right) \frac{%
\partial }{\partial \eta }-\frac{\alpha ^{2}}{1-\eta ^{2}}+R\frac{G^{2}}{4}%
Z\eta ^{2}+\Lambda _{\alpha ,R}\right\} \psi _{\beta ,\alpha }\left( R,\eta
\right) =R\lambda _{\beta ,\alpha }\left( R\right) \eta ^{2}\psi _{\beta
,\alpha }\left( R,\eta \right)  \label{2.7}
\end{eqnarray}%
$\Lambda _{\alpha ,R}$ being the separation constant. Solving the joint
eigenvalue problem (\ref{2.7}) each $R-$family of eigenstates yields the $%
\lambda _{\beta ,\alpha }\left( R\right) $ functions. However if one is only
interested in the nature of the effective $R-$potential, the problem may be
further simplified by fixing the coordinates to a fixed value. For example,
with $\xi =1$ and $\eta =0$ in the case $\alpha =0$, one obtains from (\ref%
{2.7})%
\begin{equation}
V_{\beta ,0}\left( R\right) =-G^{2}\left( \frac{4-Z}{R}\right) +\frac{4}{%
R^{2}}\psi _{\beta ,0}^{-1}\left( R,1,0\right) \left. \left( -\frac{\partial
^{2}\psi _{\beta ,0}}{\partial \eta ^{2}}-2\frac{\partial \psi _{\beta ,0}}{%
\partial \xi }\right) \right\vert _{\xi =1,\eta =0}  \label{2.8}
\end{equation}%
Therefore the effective $R-$potential contains a $1/R$ term which is
attractive for $Z<4$ and a repulsive term originating from the kinetic part
of the Hamiltonian. Because of the $1/R^{2}$ factor, this last term is
expected to be strongly repulsive at $R=0$ for the eigenstates of Eq.(\ref%
{2.5}), precluding any quasi-collisions. In fact this result is to be
expected and is a result of the factorized way the calculation is performed.
Separating the $R-$dynamics (of the mass $M$\ particles) from the dynamics
of the mass $m$ particle, what one is studying is the $R-$dynamics in the
mean field of the other particle, not the simultaneous quantum fluctuations
of all particles to the unstable or the tiny energetically favorable regions
of configuration space as described in Section $1$. Therefore to study this
phenomena, one should consider the joint dynamics of all particles. This
will be the subject of the next Section.

In the present Section prolate spheroidal coordinates were used because they
are appropriate for the factorized problem and have traditionally been used
for that purpose. However to deal with the full dynamical problem they are
not very convenient and, in addition, also not appropriate for a system
confined in a finite volume because physical space coordinates are defined
as multiples of $R$, namely%
\begin{equation*}
x=\frac{R}{2}\xi \eta ;\;y=\frac{R\sin \phi }{2}\sqrt{\left( \xi
^{2}-1\right) \left( 1-\eta ^{2}\right) };\;z=\frac{R\cos \phi }{2}\sqrt{%
\left( \xi ^{2}-1\right) \left( 1-\eta ^{2}\right) }.
\end{equation*}%
Therefore for a small $R$ the $\xi -$coordinate must be extremely large for
a physically finite volume.

\section{The dynamical problem}

\begin{figure}[htb]
\centering
\includegraphics[width=0.5\textwidth]{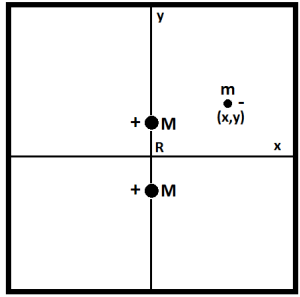}
\caption{Three particles in a
cylindrical box. A section of the box at fixed $\theta $}
\label{plane}
\end{figure}

Here one addresses the joint dynamical problem of the $3$ particles. The
reference frame is chosen with the $y-$axis along the line joining the two
mass $M$ particles and the origin at the middle point, their $y-$coordinates
being $R/2$ and $-R/2$. The three-dimensional coordinates are $\left(
x,y,\theta \right) $, $\theta $ being the angle of rotation of the plane of
the three particles. These are coordinates for a finite volume cylinder,
with the choice $x\in \lbrack -L,L]$, $y\in \lbrack -L,L]$, $\theta \in
\lbrack 0,\pi ]$ (Fig.\ref{plane}). The "radial" variable $x$ is chosen in a
symmetrical way because this is more convenient to fix the boundary
conditions.

The Hamiltonian is%
\begin{equation}
\frac{2m}{\hslash ^{2}}H=-8\mu \frac{\partial ^{2}}{\partial R^{2}}-\Delta
_{m,2}+\frac{2m}{\hslash ^{2}}V\left( R,x,y\right)  \label{3.1}
\end{equation}%
with%
\begin{equation}
\Delta _{m,2}=\frac{1}{x}\frac{\partial }{\partial x}\left( x\frac{\partial 
}{\partial x}\right) +\frac{\partial ^{2}}{\partial y^{2}}+\frac{1}{x^{2}}%
\frac{\partial ^{2}}{\partial \theta ^{2}}  \label{3.2}
\end{equation}%
\begin{equation}
V\left( R,x,y\right) =\frac{Ze^{2}}{4\pi \varepsilon _{0}}\left\{ \frac{Z}{R}%
-\frac{1}{\sqrt{x^{2}+\left( \frac{R}{2}-y\right) ^{2}}}-\frac{1}{\sqrt{%
x^{2}+\left( \frac{R}{2}+y\right) ^{2}}}\right\}  \label{3.3}
\end{equation}%
The $\theta $ dependence is taken care of by factorization with $\chi
_{\alpha }\left( \theta \right) =\exp \left( i\alpha \theta \right) $ and
two cases will be studied: first, the case of two variables $\left(
R,x\right) $, that is, the mass $m$ particle constrained to move along the $%
x-$axis and then the three variables $\left( R,x,y\right) $ case. In both
cases one considers the angular symmetric situation $\left( \alpha =0\right) 
$.

To fully grasp the nature of the dynamical problem, in the two-variables
case, both the 2-dimensional (motion in the plane) and the 3-dimensional
(motion along the $x-$axis in 3 dimensions) cases are considered. For the
three-variables case only 3-dimensional motion will be considered.

The interest of the dimensionally restricted studies is not purely academic
because, for systems confined in a molecular cage, the orbitals of the
containing molecules, that form the cage, may impose further dimensional
constraints on the confined particles.

In a confined volume the natural boundary condition to impose is the
vanishing of the wave function at the boundaries. A particularly efficient
way to obtain a very large number of eigenvalues and eigenfunctions of the
operator is a finite difference method on a grid using a fast
diagonalization routine (see for example \cite{Ogburn}). The number of
eigenfunctions that is obtained may always be improved by using finer and
finer grids. Different degrees of approximation may be used to construct the
matrix representation of the operators. In practice there is a trade-off
between using higher-order operator representations and finer grids with
lower order representations, which lead to sparser matrices. In the
following, the results obtained with a finite difference diagonalization
method are presented. Here 5-points approximations have been used for the
derivatives.

To work with dimensionless quantities one actually computes the spectrum of $%
\frac{2m}{\hslash ^{2}G^{4}}H_{2}$, the corresponding length variables being 
$G^{2}R$, $G^{2}x$ and $G^{2}y$. Recall that $G^{2}=\frac{mZe^{2}}{2\hslash
^{2}\pi \varepsilon _{0}}$.

\subsection{Two variables (R,x), $\protect\alpha =0$}

\subsubsection{Motion along the $x-$axis in the plane}

\begin{equation}
\frac{2m}{\hslash ^{2}G^{4}}H_{2}^{(P)}=-8\mu \frac{\partial ^{2}}{\partial
\left( G^{2}R\right) ^{2}}-\frac{\partial ^{2}}{\partial \left(
G^{2}x\right) ^{2}}+\left\{ \frac{Z}{G^{2}R}-\frac{2}{\sqrt{\left(
G^{2}x\right) ^{2}+\left( \frac{G^{2}R}{2}\right) ^{2}}}\right\}  \label{3.4}
\end{equation}

\begin{figure}[htb]
\centering
\includegraphics[width=0.5\textwidth]{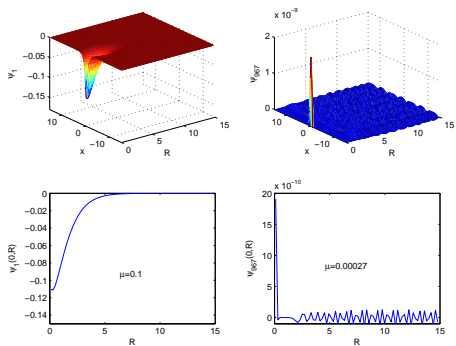}
\caption{The ground state
wave function at $\mu =0.1$ and the first quasi-collision state at $\mu =0.00027$}
\label{c0_mag0_waves}
\end{figure}

For a $150\times 150$ two-dimensional grid in the $\left( \tilde{x}=G^{2}x,%
\tilde{R}=G^{2}R\right) -$plane in a square box with $\tilde{x}$ and $\tilde{%
R}$ $\in \lbrack -15,15]$ the results are summarized in the Figs.\ref%
{c0_mag0} and \ref{c0_mag0_waves}.
\begin{figure}[htb]
\centering
\includegraphics[width=0.5\textwidth]{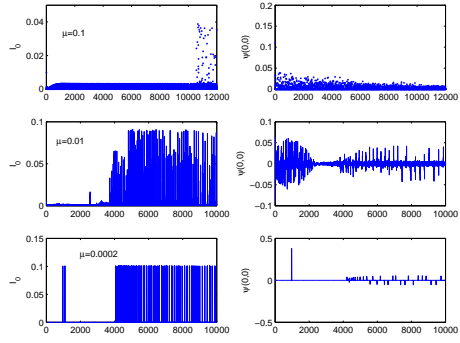}
\caption{$I_{0}$ and $\psi \left( 0,0\right) $ at several $\mu $
ratios for motion along the $x-$axis in the plane}
\label{c0_mag0}
\end{figure}

For several values of $\mu =\frac{m}{M}$ the left panels in Fig.\ref{c0_mag0}
show the value of%
\begin{equation*}
I_{0}=\int \left\vert \psi \left( \tilde{x},0\right) \right\vert ^{2}d\tilde{%
x}
\end{equation*}%
and the right panels $\psi \left( 0,0\right) $, the eigenvector value at the
origin. The number of eigenvalues listed in the figure is less than $1/2$ of
the total number of eigenvalues, higher eigenvalues being less accurate in
the finite difference method. $I_{0}$ is a measure of the quasi-collision
probability of the two heavy positively charged particles. The first such
state appears isolated high up in the spectrum, many other such states
appearing even higher in the spectrum. Notice that although all these states
have high values of $I_{0}$ they have different values at the origin $\psi
\left( 0,0\right) $. This means that although they all imply a high
quasi-collision probability $\left( R=0\right) $, they have different
spreading along the $x-$axis. The location of the first quasi-collision
state (as well as the states for which there is a non-zero value of the wave
function at the origin) move to lower energies as the $\mu $ ratio increases
and, at $\mu =0.1$ even the ground state has $I_{0}\neq 0$\footnote{%
Notice that in Fig.\ref{c0_mag0} a different plotting convention is used for 
$\mu =0.1$ (points rather than lines), to emphasize the nonzero values of $%
I_{0}$ and $\psi \left( 0,0\right) $ at the ground state.}. This means that
mass (or effective mass) of the light particle is an important
consideration. Fig.\ref{c0_mag0_waves} shows on the left panels the wave
function of the ground state for $\mu =0.1$ and in the right panels the
first quasi-collision state for $\mu =0.00027$.

Because all quantities in this paper are expressed in dimensionless
quantities all the dynamical studies in this paper may be easily adapted to
confined collisions in chemical or nuclear reactions. Actual lengths $L$ are
related to dimensionless lengths $\overset{\symbol{126}}{L}$ by $L=G^{-2}%
\overset{\symbol{126}}{L}$ and actual energies $E$ are related to the
eigenvalues $\overset{\symbol{126}}{E}$ of the dimensionless Hamiltonian $%
\frac{2m}{\hslash ^{2}G^{4}}H$ by $E=\frac{\hslash ^{2}G^{4}}{2m}\overset{%
\symbol{126}}{E}$. For definiteness, I will concentrate on the possibility
of observing nuclear reactions when nuclei are confined in solid matter.
Therefore quantitative estimates will be made for $m=$ the electron mass, $M=
$ the deuteron mass. For these values $G^{2}=$ $\frac{mZe^{2}}{2\hslash
^{2}\pi \varepsilon _{0}}=0.3779\times 10^{11}m^{-1};G^{-2}=0.2646\overset{%
\circ }{A}$ and $\tilde{x},\tilde{R}\in \lbrack -15,15]$ roughly corresponds
to 

confinement in a $8\times 8$ $\overset{\circ }{A}$ box. $\frac{m_{e}}{M_{D}}%
=0.00027$. The energy conversion factor is $\left( \frac{2m_{e}}{\hbar
^{2}G^{4}}\right) ^{-1}=54.4$ eV. Then, the binding energy of the ground
state would be $120$ eV and the first quasi-collision state would be $160$
eV above the ground state. The proliferation of the other quasi-collision
states occurs above $500$ eV. Although these results are obtained in a
simplified situation of motion along the axis on the plane, they already
indicate that spontaneous fusion of nuclei confined in solid matter either
does not occur at all or, if occurring in some random exceptional event, is
not a practical reproducible phenomenon. Consistent production of
quasi-collision states requires excitation of the system to the low $x-$ray
energy range.

The results obtained in this subsection use a basis of $21904$ states. How
the energy estimates for the quasi-collision states might depend on the
number of basis states is discussed in 3.4.

\subsubsection{\protect\bigskip Motion along the $x-$axis in the cylinder}

\begin{equation}
\frac{2m}{\hslash ^{2}G^{4}}H_{2}^{(C)}=-8\mu \frac{\partial ^{2}}{\partial
\left( G^{2}R\right) ^{2}}-\frac{1}{G^{2}x}\frac{\partial }{\partial \left(
G^{2}x\right) }\left( G^{2}x\frac{\partial }{\partial \left( G^{2}x\right) }%
\right) +\left\{ \frac{Z}{G^{2}R}-\frac{2}{\sqrt{\left( G^{2}x\right)
^{2}+\left( \frac{G^{2}R}{2}\right) ^{2}}}\right\}  \label{3.5}
\end{equation}

\begin{figure}[htb]
\centering
\includegraphics[width=0.5\textwidth]{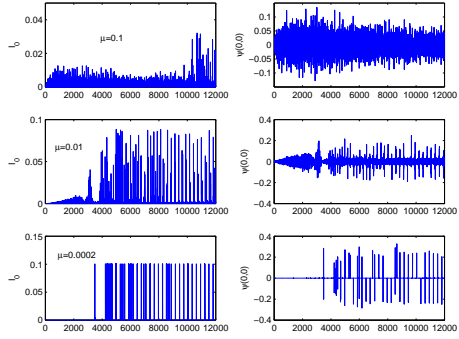}
\caption{$I_{0}$ and $\psi \left( 0,0\right) $ at several $\mu $
ratios for motion along the $x$-axis in the cylinder}
\label{c1_mag0}
\end{figure}

The situation here, as illustrated in Fig.\ref{c1_mag0}, is qualitatively
similar to the plane motion case, the main difference being that a smaller
binding energy of the ground state is obtained and the quasi-collision
states occur higher in the spectrum. With the same choices as before for the
physical parameters, the first one would be $389$ eV above the ground state
with many others above $540$ eV.

\subsection{Three variables (R,x,y), $\alpha =0$}

Here only the motion in the cylinder case will be analyzed ($\tilde{R}%
=G^{2}R $; $\tilde{x}=G^{2}x$; $\widetilde{y}=G^{2}y$)%
\begin{equation}
\frac{2m}{\hslash ^{2}G^{4}}H_{3}=-8\mu \frac{\partial ^{2}}{\partial \tilde{%
R}^{2}}-\frac{1}{\tilde{x}}\frac{\partial }{\partial \tilde{x}}\left( \tilde{%
x}\frac{\partial }{\partial \tilde{x}}\right) -\frac{\partial ^{2}}{\partial 
\widetilde{y}^{2}}+\left\{ \frac{Z}{\tilde{R}}-\frac{1}{\sqrt{\tilde{x}%
^{2}+\left( \frac{\tilde{R}}{2}-\widetilde{y}\right) ^{2}}}-\frac{1}{\sqrt{%
\tilde{x}^{2}+\left( \frac{\tilde{R}}{2}+\widetilde{y}\right) ^{2}}}\right\}
\label{3.6}
\end{equation}

\begin{figure}[htb]
\centering
\includegraphics[width=0.5\textwidth]{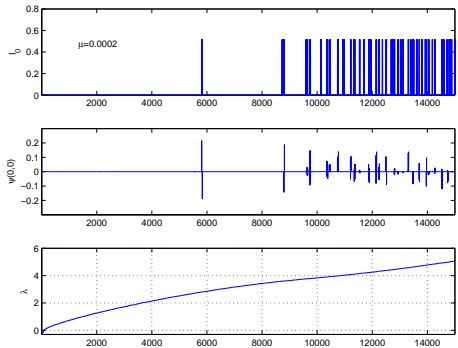}
\caption{$I_{0}$, $\psi \left( 0,0\right) $
and (dimensionless) eigenvalues for the three-variables case with $\mu =0.00027$}
\label{tripla_c1_mag0}
\end{figure}

In the previous cases, when the motion is constrained to the $y=0$ axis, the
reason why states with $I_{0}$ and $\psi \left( 0,0\right) \neq 0$ only
occur for relatively high excited states lies on the extreme narrowness of
the negative potential region when $R=0$. Then the kinetic energy of the
light particle implies an high energy contribution for localized states. In
the three variables case one would expect the existence of such localized
states to be even more energy demanding because of the instability of the
potential singularity along the $y$ direction. Nevertheless it turns out
that the spectrum situation is not very different from what it was before.
Fig.\ref{tripla_c1_mag0} (obtained with $21952$ basis states) shows the
values of $I_{0}$,%
\begin{equation*}
I_{0}=\int \left\vert \psi \left( \tilde{x},\widetilde{y},0\right)
\right\vert ^{2}d\tilde{x}d\widetilde{y}
\end{equation*}%
$\psi \left( 0,0,0\right) $ and the corresponding dimensionless eigenvalue
values $\lambda =\overset{\symbol{126}}{E}$, for $\mu =0.00027$.

The first quasi-collision state occurs at $\widetilde{E}=2.8$ which would
correspond to $152$ eV with many others after $3.59$ ($195$ eV). Excitation
of these states are as before in the low $x-$ray energy range. Notice
however that these values should only be considered as lower bounds for the
excitation energies, because a smaller spatial density of basis states has
been used, as compared to the one in the two previous subsections (see the
discussion in 3.4).

\subsection{Confinement and basis size effects}

An important issue is the dependence of the quasi-collision states on the
size of the confinement box and on the number of basis states that is used
to compute the spectrum.

Concerning the dependence on the size of the confinement box it is found
that the energies of the quasi-collision states grow when the size of the
box decreases, but they appear earlier in the spectrum.\begin{figure}[htb]
\centering
\includegraphics[width=0.5\textwidth]{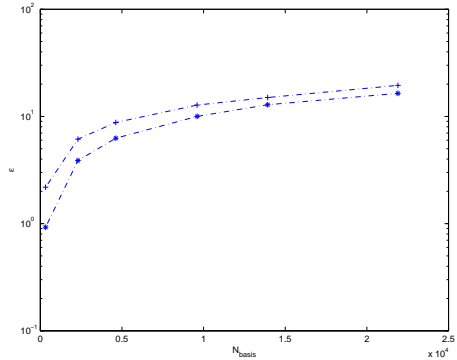}
\caption{Dependence on the number of basis states of
the energy of the first quasi-collision state (lower line) and the energy above which many other such states exist (upper line)}
\label{states}
\end{figure}

Of more interest for the correct estimation of the energy needed to excite
the quasi-collision states is their dependence on the number of states used
for the computation of the spectrum. Fig.\ref{states} shows the dependence
on the number of basis states of the energy of the first quasi-collision
state (lower line) and the energy above which many other such states exist
(upper line). This calculation refers to the two-variables in the cylinder
case of subsection 3.2. Whereas the ground state energy is found to be quite
stable for sizes above 800, the quasi-collision energies still grow even at
sizes above 20000. Therefore the values obtained in the previous subsections
should be considered as lower bounds for the correct excitation energies,
which however, from the behavior seen in the figure, should still be
expected to be in the $x-$ray range for the physical parameters used here.

\section{Quantum control of quasi-collisions}

From the dynamical study of the previous section one sees that, in spite of
the strong Coulomb barrier there indeed are \textit{"quantum quasi-collision
states"} (in the sense of the definition before, $I_{0}\neq 0$) of the three
particles. They are located at energies well above the ground state, the
question being whether the system may be driven to these states by practical
means. This is a typical question of quantum control. At the present day the
only viable way to quantum control is through the electric field of laser
pulses with, eventually, as will be seen later, a tuning effect of magnetic
fields.

In the dipole approximation the Hamiltonian for particles interacting with
the electromagnetic wave of a laser pulse is%
\begin{equation}
i\hslash \frac{\partial }{\partial t}\psi \left( x_{j},t\right)
=\sum_{j}\left\{ H_{0}-e_{j}x_{j}\cdot E\left( t\right) \right\} \psi \left(
x_{j},t\right)  \label{4.1}
\end{equation}%
with%
\begin{equation}
H_{0}=\sum_{j}-\frac{\hslash ^{2}}{2m_{j}}\nabla _{j}^{2}+V\left(
x_{j}\right)  \label{4.2}
\end{equation}%
$V\left( x_{j}\right) $ being the Coulomb interactions in (\ref{3.4} - \ref%
{3.6}) or these interactions complemented by an external magnetic field, as
described later. Let the eigenstates of $H_{0}$ be known

\begin{equation}
H_{0}\phi _{k}=\varepsilon _{k}\phi _{k}  \label{4.3}
\end{equation}%
The goal is, starting from an initial state $\phi _{i}$ (typically the
ground state), to lead the system to a desired final state $\phi _{f}$ (here
a quasi-collision state). Treating the term $H_{I}=-\sum_{j}e_{j}x_{j}\cdot
E\left( t\right) =-\sum_{j}e_{j}E\exp \left( i\omega t\right) $ as a
perturbation, the evolution operator $U_{I}\left( T,0\right) $ in the
interaction picture is%
\begin{equation}
U_{I}\left( T,0\right) =1-\frac{i}{\hslash }\int_{0}^{T}H_{I}\left(
t^{\prime }\right) U_{I}\left( t^{\prime },0\right) dt^{\prime }  \label{4.4}
\end{equation}%
with%
\begin{equation}
H_{I}\left( t\right) =-\sum_{j}e_{j}Ee^{i\omega t}e^{\frac{i}{\hslash }%
H_{0}t}x_{j}e^{-\frac{i}{\hslash }H_{0}t}  \label{4.5}
\end{equation}%
\begin{equation}
e^{\frac{i}{\hslash }H_{0}t}x_{j}e^{-\frac{i}{\hslash }H_{0}t}=x_{j}-it\frac{%
\hslash }{m_{j}}x_{j}\frac{\partial }{\partial x_{j}}+t^{2}\frac{\hslash }{%
4m^{2}}V^{\prime }\left( x_{j}\right) +\cdots   \label{4.6}
\end{equation}%
The transition probability from the initial state $\phi _{i}$ to the final
state $\phi _{f}$ is obtained from $\left\vert \left\langle \phi
_{f}\left\vert U\left( T,0\right) \right\vert \phi _{i}\right\rangle
\right\vert ^{2}$. When the desired final state is a quasi-collision one,
the contribution of the $U\left( T,0\right) $ series will be strongly
suppressed by the $x_{j}$ terms, which vanish at the collision. Therefore
one expects the leading contribution to be%
\begin{equation}
\left\vert \frac{\hslash E}{4m^{2}}\int_{0}^{T}e^{-\frac{i}{\hslash }t\Delta
\varepsilon }e^{it\omega }t^{2}\left\langle \phi _{f}\left\vert V^{^{\prime
}}\left( x_{j}\right) \right\vert \phi _{i}\right\rangle \right\vert 
\label{4.7}
\end{equation}%
$\Delta \varepsilon $ being the energy difference between $\phi _{f}$ and $%
\phi _{i}$. From (\ref{4.7}) one concludes that in addition to a laser
frequency tuned to $\Delta \varepsilon $, long duration pulses should be
favored.

\subsection{Spectrum modulation by a constant magnetic field}

Here, one analyses the shifts in the spectra studied in Section 3 which
might be obtained with an external static (or slowly varying) magnetic
field. With an external field the electromagnetic contribution to the
Hamiltonian is%
\begin{equation}
H_{em}=\frac{1}{2m}\left( p-eA\left( x,t\right) \right) ^{2}+e\Phi \left(
x,t\right)   \label{5.1}
\end{equation}%
which in the Coulomb gauge, $\nabla \cdot A=\Phi =0$, becomes%
\begin{equation}
H_{em}=-\frac{\hslash ^{2}}{2m}\nabla ^{2}+i\frac{e\hslash }{m}A\cdot \nabla
+\frac{e^{2}}{2m}A^{2}  \label{5.2}
\end{equation}%
The two interaction terms are of a different nature, the first one being
called the paramagnetic \ term and the last the diamagnetic term. For a
stationary uniform magnetic field, let%
\begin{equation}
A\left( x\right) =-\frac{1}{2}x\times B=-\frac{1}{2}\left\{ e_{x}\left(
yB_{z}-zB_{y}\right) +e_{y}\left( zB_{x}-xB_{z}\right) +e_{z}\left(
xB_{y}-yB_{x}\right) \right\}   \label{5.3}
\end{equation}%
Then the paramagnetic term is%
\begin{equation}
i\frac{e\hslash }{m}A\cdot \nabla =-\frac{e}{2m}L\cdot B  \label{5.4}
\end{equation}%
with%
\begin{equation}
L=x\times \left( -i\hslash \nabla \right)   \label{5.5}
\end{equation}%
For the simplest cases studied in Section 3, $L=0$, the paramagnetic term
term vanishes, the only contribution coming from the diamagnetic term. The
contribution of the diamagnetic term to the dimensionless Hamiltonians $%
\frac{2m}{\hslash ^{2}G^{4}}H$ is 
\begin{equation}
\frac{e^{2}}{\hslash ^{2}G^{8}}\left( \frac{1}{4}\widetilde{x}^{2}\left(
B_{z}^{2}+B_{y}^{2}\right) +\frac{Z^{2}\mu }{4}\widetilde{R}^{2}\left(
B_{z}^{2}+B_{x}^{2}\right) \right)   \label{5.6}
\end{equation}%
This adds to the dynamics an harmonic contribution which would favor a
closer proximity of the particles. Notice however that $\frac{e}{\hslash
G^{4}}$ is the factor which multiplies the physical fields which, for the
physical parameters used before (three-body quasi-collision of deuterons),
is extremely small, of order $1.06\times 10^{-6}$. Therefore the
contribution of the diamagnetic interaction term would be too small to be of
practical relevance in this case. It might however be relevant for other
physical parameters, namely chemical reactions of confined atoms.

By contrast the corresponding factor in the paramagnetic interaction term
would be $\frac{e}{\hslash ^{2}G^{4}}$ and physically reasonable magnetic
fields may induce appreciable spectrum shifts in the $L\neq 0$ case.

\section{Remarks and conclusions}

1 - Properties of confined systems may greatly differ from similar systems
in free spaces. Exploration of new pathways for chemical or nuclear
reactions might be a promising application for atoms or nuclei confined in
molecular cages.

2 - In what concerns the possibility to observe fusion reactions by
many-body effects when nuclei are confined in a cage, the main conclusion of
this paper is that spontaneous occurrence of these events is quite
improbable and if they occur at all under ergodic situations they will be
basically uncontrollable and irreproducible. Nevertheless, considering the
molecular cage merely as a confinement device, reproducible quasi-collisions
might be induced by quantum control techniques. This two-step protocol would
be what elsewhere \cite{Vilela4} has been called "hybrid fusion".

3 - Quantum control is a technique that has had in recent years remarkable
development. Learning and adaptive techniques \cite{Shapiro} \cite{Rabitz},
optimal control \cite{Frank}, unitary and non-unitary \cite{Vilela5} \cite%
{Mandilara} evolution methods have been developed, even infinite-dimensional
spaces once considered to be uncontrollable have been proved to yield to
full quantum control \cite{Karwowski} \cite{Vilela6}. Here only a very basic
discussion of the control requirements has been performed. Once the detailed
nature of the molecular cage is specified, all the developed techniques may
be applied to smoothly drive the system to the quasi-collision states.

4 - For the fusion situation that was specified, excitation of the
quasi-collision states seemed to require laser pulses on the $x-$ray range.
It is interesting to note that also on a recent experiment some authors \cite%
{Belyaev} suggest the induction of fusion reactions in a crystal by $x-$rays.

5 - A point that should be recalled when identifying fusion reactions
induced by many-body effects is that the reaction channels and final
products might be different from those of the two-body reaction \cite%
{Vilela-pre} \cite{Vilela3}.

6 - Here the problem of three-particles in a single molecular cage was
considered and quantum control of elementary quasi-collision states has been
emphasized. Another situation where similar quasi-collisions might occur is
when many such contiguous cages communicate and the collective system is
sufficiently excited to be describable by a chaotic ergodic measure. This is
the situation studied in \cite{Vilela3}, where small but non-negligible
quasi-collisions rates were found. However, the chaotic nature of the
collective events would render them either difficult to control or
irreproducible and therefore of little interest for steady energy production
applications. This situation might however be relevant as a correction to
the dynamics of stellar models.

\end{document}